\documentclass[%
preprint,
 amsmath,amssymb,
 aps,nofootinbib,
]{revtex4-1}
\usepackage{euscript,amsmath,amssymb,amsfonts,graphicx,bm}

\usepackage[english]{babel}
\usepackage{amsthm}
\usepackage{paralist}
\usepackage{epstopdf}
\usepackage{enumitem}
\usepackage[colorlinks=true]{hyperref}
\hypersetup{urlcolor=black, citecolor=black}
\usepackage{xcolor}

\allowdisplaybreaks
\usepackage{xcolor}
 
\usepackage[utf8x]{inputenc}

\usepackage{textcomp,marvosym}

\DeclareMathOperator*{\argmax}{argmax} 

\date{}

\usepackage{lastpage,fancyhdr,graphicx}
\usepackage{epstopdf}

\setlist[itemize]{leftmargin=*}

\begin{document}
\vspace*{0.2in}

\begin{flushleft}
{\Large \bf Stochastic dynamics of social patch foraging decisions}
\newline
\\
\bigskip
Subekshya Bidari\textsuperscript{1,*}, Ahmed El Hady\textsuperscript{2, 3, 4,$\dagger$}, Jacob Davidson\textsuperscript{3,$\ddagger$} and Zachary P Kilpatrick\textsuperscript{1,**}
\\
\bigskip
\textbf{1} Department of Applied Mathematics, University of Colorado, Boulder CO, USA \\
\textbf{2} Princeton Neuroscience Institute, Princeton, New Jersey, USA \\
\textbf{3} Department Collective Behavior, Max Planck Institute for Animal Behavior, Konstanz, Germany
\\
\textbf{4} Cluster for Advanced Study of Collective Behavior, Konstanz, Germany
\\
\textbf{*} subekshya.bidari@colorado.edu \\
\textbf{$\dagger$} ahady@princeton.edu \\
\textbf{$\ddagger$} jdavidson@ab.mpg.de \\
\textbf{**} zpkilpat@colorado.edu
\end{flushleft}

\section*{Abstract}
Animals typically forage in groups. Social foraging can help animals avoid predation and decrease their uncertainty about the richness of food resources. Despite this, theoretical mechanistic models of patch foraging have overwhelmingly focused on the behavior of single foragers. In this study, we develop a mechanistic model that accounts for the behavior of individuals foraging together and departing food patches following an evidence accumulation process. Each individual's belief about patch quality is represented by a stochastically accumulating variable which is coupled to others' belief to represent the transfer of information. We consider a cohesive group, and model information sharing by considering both intermittent pulsatile coupling (only communicate decision to leave) and continuous diffusive coupling (communicate throughout the evidence accumulation process). We find that foraging efficiency under pulsatile coupling has a stronger dependence on the coupling strength parameter compared to diffusive. Despite employing minimal information transfer, pulsatile coupling can still provide similar or higher foraging efficiency compared to diffusive coupling. Conversely, since diffusive coupling is more robust to parameter choices, it performs better when individuals have heterogeneous departure criteria and social information weighting. Efficiency is measured by a reward rate function that balances the amount of energy accumulated against the time spent in a patch, computed by solving an ordered first passage time problem for the patch departures of each individual. Using synthetic data we show that we can distinguish between the two modes of communication and identify the model parameters.  Our model establishes a social patch foraging framework to parse and identify deliberative decision strategies, to distinguish different forms of social communication, and to allow model fitting to real world animal behavior data.


\noindent
{\bf Keywords:} patch-leaving decisions, foraging, drift-diffusion model, first passage times


\section{Introduction} 
Foraging is a ubiquitous behavior performed by all animals for survival, and many species forage in groups. 
Communication among group members is essential to maintain cohesion and share important information, e.g.\ regarding available resource or possible threats~\cite{conradt2005consensus,miller2013both}.
In general, the advantages of foraging in groups may include increased vigilance and protection against predators \cite{clark1986evolutionary,powell1974experimental,siegfried1975flocking}, faster estimates and minimization of variance in patchy resources or resources of uncertain quality \cite{krebs1972flocking, ward1973importance}, and  enhanced ability to capture prey \cite{dumke2018advantages}. 
Since resource distributions can be spatially heterogeneous \cite{weimerskirch2007seabirds,fauchald1999foraging,levin2000multiple}, a common modeling assumption takes resources disbursed in patches \cite{levin1976population,mcmahon2017habitat,mitchell2004mechanistic} with little availability in between. In the framework of patch foraging, animals exploit resources in a patch until departing for another patch. A key question is then the process and strategy that determines an animal's patch departure decision. Although many animals forage in groups and use social information to shape their movement and resource exploitation decisions, quantitative models have mainly focused on mechanistic models of individual foragers~\cite{davidson19,kilpatrick21}. In this study, we introduce ``social patch foraging" to describe the behavior involving two or more animals foraging in a patchy environment and ask how the social information modulates the process preceding individual foragers' patch departures.


Animal groups such as baboons or capuchins exchange information and coordinate their movement to remain mostly as a cohesive group~\cite{strandburg2015shared,strandburg2017habitat}. Bayesian models have proposed optimal ways to combine individual and social information in order to gain information about the environment~\cite{perez2011collective}. Theory has been used to show how group structure and communication affect behavior, determining when differences in information drive a group to split apart, or the fraction of informed individuals needed to lead a group to a known location \cite{sueur2011group,couzin2005effective}. Other studies have used models of contagion \cite{dodds2004universal,pagliara2020adaptive} to examine how a behavior spreads through a group \cite{rosenthal2015revealing}. One observed advantage of information sharing in groups is that multiple estimates of the same quantity (e.g., chemical gradients or food density) reduce uncertainty arising from measurement or internal noise~\cite{srivastava2014collective,ellison2016cell}. While uncertain and/or noisy decision processes ultimately limit patch foraging efficiency~\cite{kilpatrick21}, such effects may be ameliorated by social communication.

The stochastic departure time of each individual in patch leaving models can be calculated from an accumulation to bound process, wherein a drifting and diffusing variable represents a current belief and a fixed bound triggers a patch departure~\cite{davidson19}. Such drift-diffusion models have been successful in untangling the strategies animals use to make binary perceptual choices~\cite{gold2007neural}. Decision times can then be obtained as the solutions to first passage time problems using a variety of techniques from asymptotic methods and stochastic processes~\cite{bogacz2006physics,roldan2015decision,gardiner09}. The behavior of such stochastic bound-crossing problems are relevant not only to understanding decision making but also neural spiking~\cite{lindner2002maximizing}, search processes~\cite{benichou2014first}, and biomolecular trafficking~\cite{schuss2007narrow}. 

%
Leveraging the drift-diffusion modeling framework, we formulate a social patch foraging model and study how information sharing in animal groups and heterogeneity of beliefs affect patch-leaving decisions and the efficiency of group decisions.
In the model, we consider a cohesive foraging group, and two different ways individuals might share their evolving beliefs - diffusive or pulsatile coupling. Diffusive coupling represents continuous sharing of information, and this has previously been used to describe humans who share their decision processes (an “ideal group” for group decision-making \cite{sorkin2001signal,srivastava2014collective}), as well as groups of migrating animals \cite{pais2014adaptive,torney2010specialization}. Pulsatile coupling involves only sharing information when a decision is made, as discussed in \cite{caginalp17,karamched20}.
In general, our results suggest that increasing the coupling strength between individuals improves the efficiency of group decisions (i.e., increasing the average rate of energy intake) through more coordinated individual departure decisions. For the cohesive foraging groups we consider in the model, we find that precise alignment in patch departure is more important for group efficiency than the tuning of individual departure times - this is because when individual decision times differ, one agent must wait for another after they stop foraging in the patch, and this waiting time reduces the overall time spent foraging. We find by detuning parameters of both models that the diffusively coupled model is more robust to suboptimal parameter choices. Following this, we develop model-fitting methods and determine identifiability using synthetic data (i.e., patch departure times generated by the models). This provides a clear framework for fitting proposed models to experimental field data and inferring modes of communications used by animals in natural social patch foraging decisions.

\section{Model and Methods}
\begin{figure}[t!]
\begin{center}  \includegraphics[width=17cm]{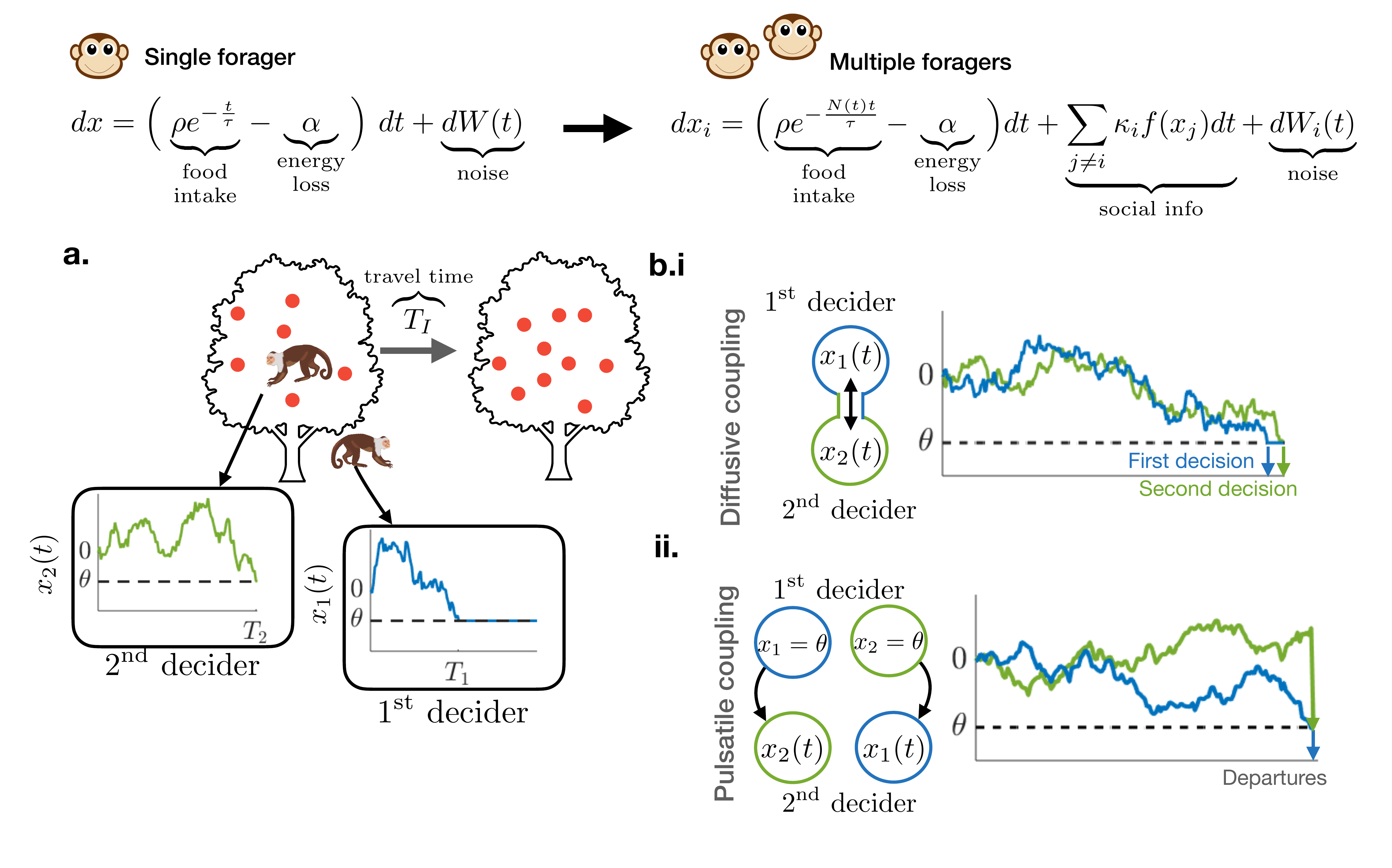}	\end{center} 
\vspace{-25pt}
\caption{Evidence accumulation model for patch leaving in individuals and groups. {\bf a.} Schematic of patch departure decision in a group with two agents  {\bf b.i.} \emph{Diffusive coupling} - agents communicate their beliefs throughout the evidence accumulation process to their neighbors; {\bf b.ii.} \emph{Pulsatile coupling} - agents only communicate their decision to leave the patch to their neighbors.}
\label{fig1}
\end{figure}
Our stochastic model describes agents' beliefs about the quality of the patch where they are currently foraging. We consider a cohesive group where all individuals in the group leave the patch together once all group members have made a departure decision. Thus, there are two key time points for any individual: when that individual $i$ decides to leave and stops foraging $(T_i)$, and when the whole group leaves ($T_N$).
The group leaving time is the maximum individual decision time among the $N$ group members ($T_N = \max(\{T_i\})$), and therefore a high variance in $T_i$ leads to a low average returns, due to the time early individuals spend waiting. To illustrate this process with a concrete example, consider capuchin monkeys foraging on fruit trees. The time $T_i$ is when individual $i$ comes down from the tree, and the time $T_N$ is when all individuals have come down from the tree and the group leaves together. 


Prior to introducing models with information coupling, we discuss a single forager model where the belief $x$ of a foraging agent is driven by the noisy sampling of a depleting resource in a patch, which starts at $\rho$ and decays with timescale $\tau$ as the agent forages. The evidence accumulation process of a single forager evolves according to the following equation
\begin{align}
d x = ( \rho e^{-\frac{t}{\tau}} - \alpha) \ dt + \sqrt{2 B} \ dW(t), \ \ \ x(0) = 0, \ \ \ x(T) = \theta,	\label{expsde}
\end{align}
where $\alpha$ is the cost associated with foraging, and $W(t)$ is the standard Wiener process (zero mean, variance unity). The agent will depart the patch at a stochastically determined time $t=T$ such that $x(T) = \theta < 0$. A similar model of an individual foraging in a patchy resource and individual decision strategies was previously analyzed for its optimal and near-optimal regimes~\cite{davidson19}.

We incorporate information transfer among individuals to represent foraging social groups. 
Information sharing can take the form of social cues and signals (inadvertent or intentional) emitted to influence the behavior of conspecifics \cite{king2011next,black1988preflight,stewart1994gorillas,boinski1995use,poole1988social} and is mathematically represented in our models as the coupling of decision states between individuals in a group~\cite{caginalp17,karamched20}. We consider two different information-sharing mechanisms: \\[0.5ex]
(i) {\em Diffusive coupling:} Agents communicate their beliefs continuously throughout the evidence accumulation process, and are attracted to the relative beliefs of their neighbors according to an individual coupling strength $\kappa_{j}$. A similar formulation has been considered previously in the context of drift-diffusion models with constant drift~\cite{srivastava2014collective}. The corresponding stochastic differential equations evolve as
\begin{align}    \label{eqdiff} 
d x_i =& \left( \rho e^{-\frac{N(t) \, t}{\tau}} - \alpha \right) \ dt  + \sum_{j \neq i} \kappa_i ((x_j - \theta_j) - (x_i - \theta_i))\ dt + \sqrt{2 B} \ dW_i(t), \\
x_i(0) =& 0, \ \ \ x_i(T_j) = \theta_i.
\end{align} 
Note that the decision threshold is $\theta$, and therefore $(x_i-\theta_i)$ is the distance-to-threshold for agent $i$ ($\theta<0$, so that $x_i>\theta_i$ until reaching threshold). In this formulation, an agent compares their distance-to-threshold with other agents. If other agents $j \neq i$ have not yet reached threshold, this yields a positive drift so that agent $i$ will remain longer in the patch. If all agents $j \neq i$ have already reached threshold, then $x_i$ drifts at maximum rate towards the decision threshold. Through this process, the coupling therefore tends to synchronize the decision variables and departure times of agents.

For individuals foraging in a group, the resource decay rate increases proportionally to the number of agents $N$ in a patch. We represent this with $N(t)$, which is the decreasing counting function
\begin{align}   \label{Nfun}
    N(t) = \begin{cases} N \quad & 0<t<T_1 \\ N-1 & T_1<t<T_2 \\ \vdots \\ 1 & T_{N-1}<t<T_N \end{cases},
\end{align}
where $T_j$ is the time when the $j^{\text{th}}$ decider reaches their decision threshold ($x_i(T_j) = \theta_i$), after which their belief state remains there. This continues until the final agent  has decided when $x_i(T_N) = \theta_i$. \\[0.5ex]
\noindent
(ii) {\em Pulsatile coupling:} Agents only communicate a pulse of information to their neighbors when they decide to stop foraging. When an agent's belief state reaches the threshold, neighboring agents receive a pulse in their belief state towards the threshold corresponding to their coupling strength $\kappa_{i}$. The belief states of agents evolve as
\begin{align}  \label{eqpulse} 
d x_i =& \left( \rho e^{-\frac{N(t) \, t}{\tau}} - \alpha \right) \ dt - \sum_{j \neq i} \kappa_i \delta (x_j - \theta_j) \ dt  + \sqrt{2 B} \ dW_i(t), \\
x_i(0) =& 0, \ \ \ x_i(T_j) = \theta_i,
\end{align}
where $N(t)$ is defined in Eq.~(\ref{Nfun}) and $\delta(x)$ is the delta distribution. When the first agent decides to leave (at time $t = T_1$), undecided agents ($x_j$) receive a pulse of size $\kappa_j$, propelling their belief towards their decision threshold. This may immediately trigger remaining agents to make a decision ($x_j(T_1^+) \leq \theta_j$), or the agents may continue to accumulate evidence until reaching threshold $\theta_j$.

\section{Results}
Collectives can improve their foraging efficiency by coordinating spatial movements and patch  departures so the  group remains cohesive~\cite{conradt2005consensus}. To identify communication mechanisms that could underlie information sharing and group coherence, we compare two different communication strategies, as described above. Individuals may continually adjust their beliefs to align with their group mates' during the patch foraging process (diffusive) or may only observe and incorporate information about when their neighbors decide to leave a patch and stop foraging (pulsatile). Each agent's belief evolves according to their privately collected information and the social information (received from their neighbors), until the ``decision time" when this belief reaches threshold. Each agent stops foraging when its belief reaches threshold. The group departs the patch when the last decider reaches threshold. Thus, we distinguish between individual decision times (when each agent reaches their respective decision threshold and stops foraging) and departure time (when the group leaves the patch together).

Efficiency of the group's decision is measured according to the average reward rate for all group members, averaged across many patch bouts. Assuming that the rate of resource consumption by the agents equals the rate of resource decay, 
\begin{align}
    \frac{dr}{dt} = \rho e^{- \frac{N(t) \cdot t}{\tau}}
\end{align} 
the total amount of food taken in by time $T = T_N$ is
\begin{equation} \begin{split}
    r(T) = \rho \tau \Big[ \left( 1- e^{- \frac{N \cdot T_1}{\tau}} \right) + e^{- \frac{N \cdot T_1}{\tau}}\left(1- e^{- \frac{(N-1) (T_2-T_1)}{\tau}} \right) + \cdots \\+ e^{- \frac{N \cdot T_1}{\tau}} e^{- \frac{(N-1) (T_2-T_1)}{\tau}} \cdots \left(1- e^{- \frac{(T_N-T_{N-1})}{\tau}} \right) \Big].
\end{split} \end{equation}  
Accounting for the travel time between patches $T_I$ and a constant energy loss rate $\alpha$ due to movement, we define the reward rate in the patch as 
\begin{align} \label{rewardeq}
    RR = \frac{ \langle r(T) \rangle - \alpha (T_I + \langle T \rangle)}{T_I + \langle T \rangle},
\end{align}
where the reward and departure time are averaged across realizations for a given strategy. Efficiency of a group foraging strategy is thus measured by the relative value of RR. We compute the expected reward rate in the patch by averaging the distributions of patch residence times computed from the first passage times of the threshold crossing processes defined by Eq.~(\ref{eqdiff}) and Eq.~(\ref{eqpulse}).

Before delving into the optimality of different coupling mechanisms, we compare limiting cases in a symmetric two agent system where agents have the same decision threshold $\theta_j= \theta$ and coupling constant $\kappa_j = \kappa$ for $j=1,2$ that is either absent ($\kappa = 0$) or infinite ($\kappa \to \infty$).

\subsection{Perfectly coupled homogeneous two agents system} \label{perfcoup}
\begin{figure}
\begin{center}  \includegraphics[width=15cm]{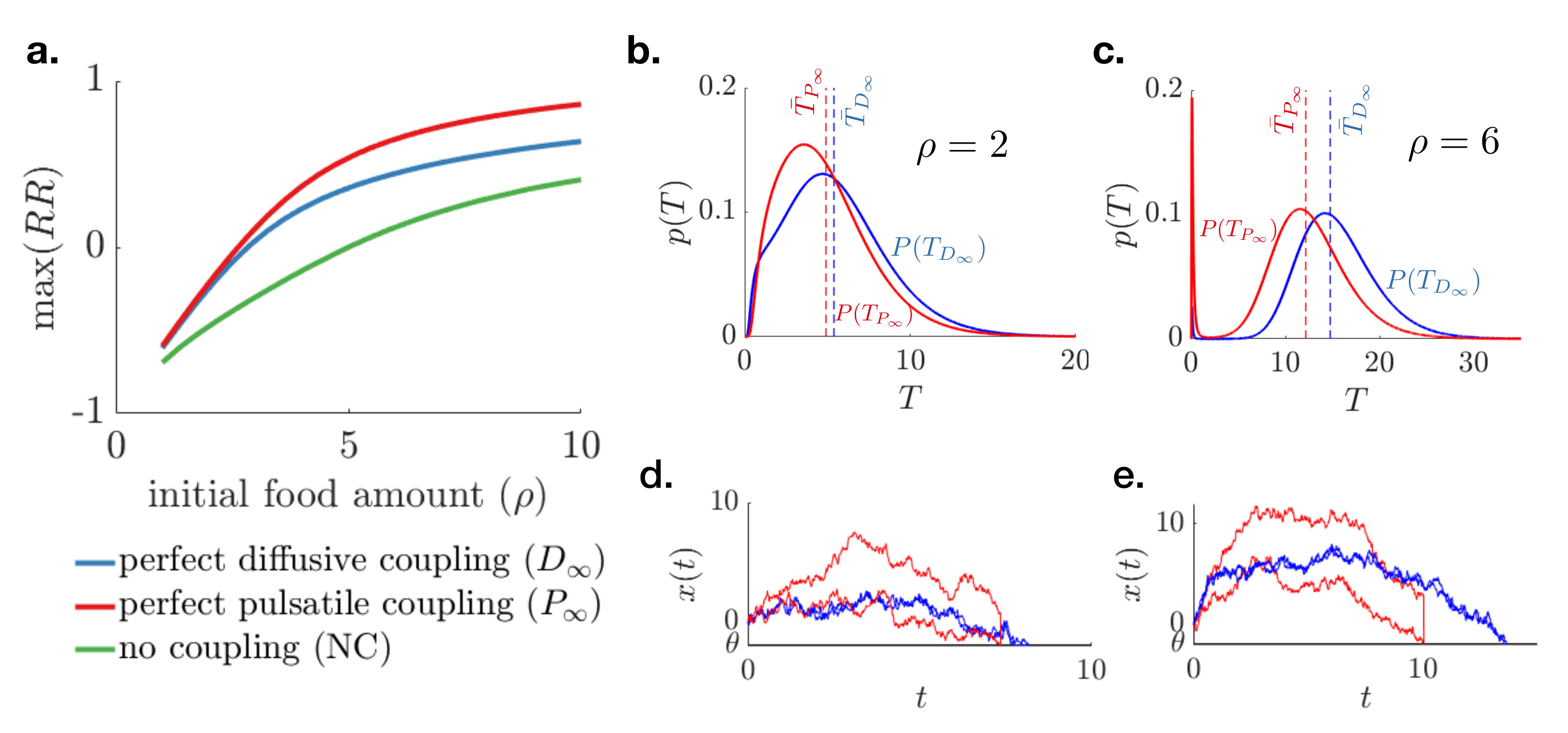}	\end{center} 	
\vspace{-25pt}
\caption{Coupling of belief states between group members increases the reward rate ($RR$). {\bf a.}~Comparison of reward rate for a two agent system: perfectly diffusively coupled ($D_{\infty}$), perfectly pulsatile coupled ($P_{\infty}$), and no information coupling (NC).  For lower values of initial food, diffusive and pulsatile coupling perform similarly generating higher reward compared to the uncoupled case. When the initial food amount is higher, pulsatile coupling outperforms diffusive coupling. Model parameters fixed at $\tau = 5, B = 1$, and $\alpha = 1.$ {\bf b,c.} Patch residence time distribution for diffusive (blue) and pulsatile (red) coupling. The dashed line shows average patch residence time. A single simulation of pulsatile (red) and diffusive coupled (blue) models when {\bf d.} $\rho = 2$ {\bf e.} $\rho = 6$.}
\label{fig2}
\end{figure}
We start by comparing groups' patch residence times ($T_N$) and reward rates (RR) for various idealized conditions, including (1) No information coupling (NC), (2) Perfect diffusive coupling ($D_\infty$), and (3) Perfect pulsatile coupling ($P_{\infty}$).

With no information coupling (NC), the decision variable for each agent evolves as an independent and identically distributed (i.i.d) process
\begin{align} \label{nceqn}
    d x_i = \left( \rho e^{-\frac{N(t) \, t}{\tau}} - \alpha \right) \ dt  + \sqrt{2B} \ dW_i(t), \ \ \ x_i(0) = 0, \ \ \ x_i(T_j) = \theta,
\end{align} 
where $N(t)$ is a decreasing counting function as in Eq.~(\ref{Nfun}).

In a perfect diffusively coupled model ($D_{\infty}$), each agent's belief evolves identically due to the averaging of all agents' drift and diffusion terms, reducing Eq.~(\ref{eqdiff}) to
\begin{align}   \label{asmpdiff}
    d x = \left( \rho e^{-\frac{2 \, t}{\tau}} - \alpha \right) \ dt  + \sqrt{B} \ dW(t), \ \ \ x(0) = 0, \ \ \ x(T) = \theta.
\end{align} 
Note, the noise amplitude of this averaged equation is half that of the full model in Eq.~(\ref{eqdiff}) due to the noise-cancellation effects of diffusive coupling~\cite{gardiner09}.  Since the agents' decision times are identical, the group's patch departure time is given by the first passage time of Eq.~(\ref{asmpdiff}). 

In a model with perfect-pulsatile coupling ($P_{\infty}$), the first decider always immediately triggers departure decisions from the remainder of the group. Thus, group decision time is given by the minimum first passage time of $j=1,...,N$ of
\begin{align} \label{asmpulse}
    d x_i = \left( \rho e^{-\frac{2 \, t}{\tau}} - \alpha \right) \ dt  + \sqrt{2B} \ dW_i(t), \ \ \ \ x_i(0) = 0, \ \ \ \ x_i(T_j) = \theta.
\end{align}

Compared to the no-coupling case, having information coupling increases the group reward rate (Fig.~\ref{fig2}a).
From the plotted reward rates, we see that for this limiting case of perfect coupling, whether diffusive or pulsatile coupling leads to a higher reward rate depends on the parameter regime. With infinitely strong coupling, the diffusive models has lower effective noise, but in both models, agents make departure decisions at the same time. For high $\rho$ values (Fig.~\ref{fig2}a), increasing the noise amplitude increases the overall average RR, by lowering the average decision time (Fig.~\ref{fig2}b-e). Note the sharp peak in the likelihood of early departure times for the pulsatile coupled model given more initial food (Fig.~\ref{fig2}c). This is because the optimal threshold $\theta^{\rm opt}$ sits closer to the initial belief at $x_i(0) = 0$. These results demonstrate that while social coupling generally increases the average reward rate for a cohesive group, the results depend on the parameter regime. While the drift and threshold values can be tuned to maximize RR for a solitary individual or for the idealized coupling case \cite{davidson19}, here we fix the drift, and focus on a comparison of the two coupling methods and their parameter dependence in a finite range.


\subsection{Decisions in diffusive coupled system}
\begin{figure}[h!]
\begin{center}  \includegraphics[width=14cm]{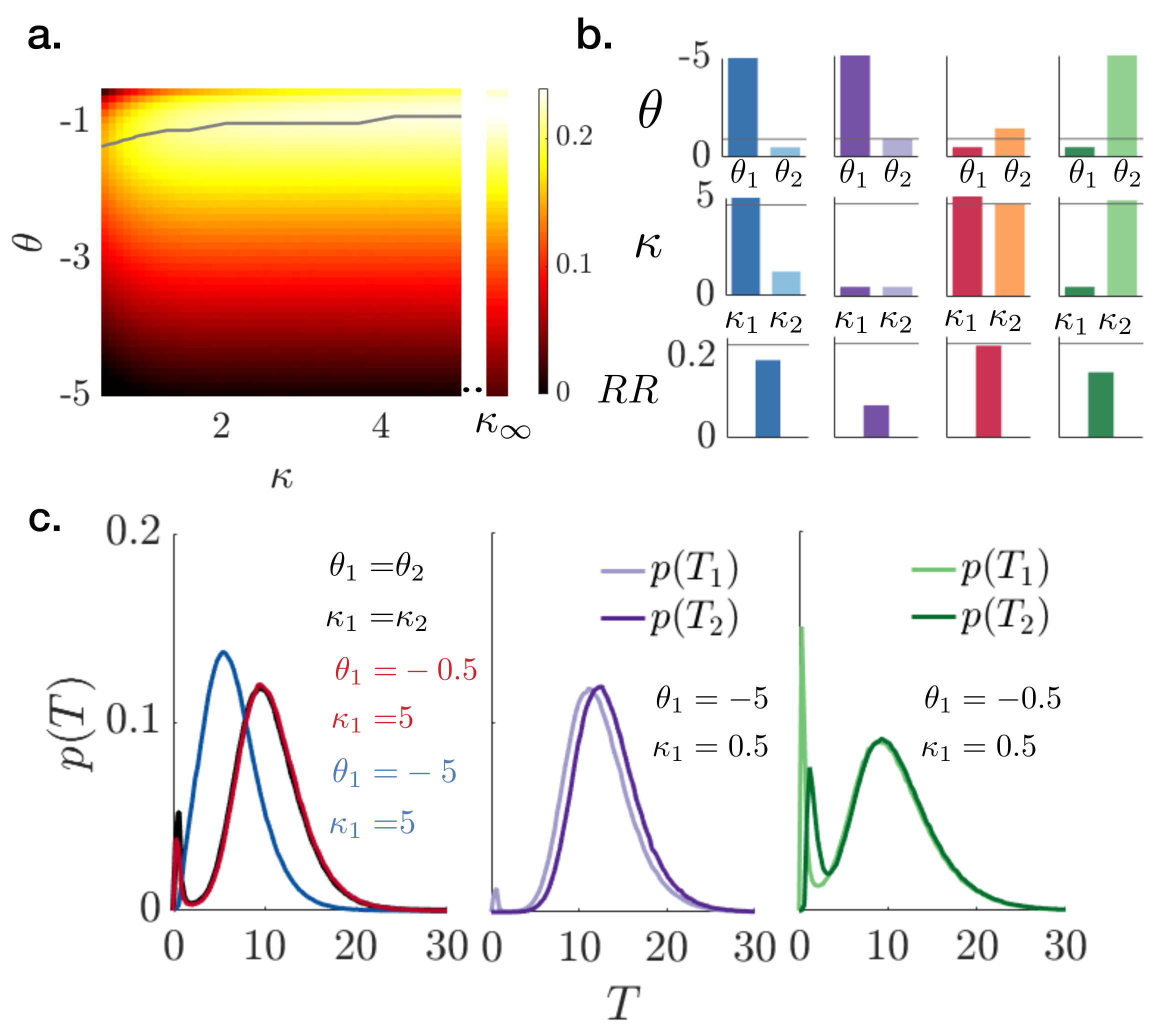}	\end{center} 
\vspace{-25pt}
\caption{Comparison of reward rates (RRs) in a group with diffusively coupled belief states. {\bf a.}~Heat map shows how reward rate varies with the decision threshold $\theta$ and coupling strength $\kappa$ for a symmetric model $\theta_1 = \theta_2 = \theta$ and $\kappa_1 = \kappa_2 = \kappa$, and $\alpha = 1, \rho=4, \tau=5$. Grey line: Optimal threshold value $\theta^{\rm opt}$ for fixed coupling strength. The line is smoothed by computing averages over a sliding window of length $3$. Right column: RR for the infinite coupling limit ($\kappa \to \infty$). {\bf b.}~Asymmetric strategies. Optimal threshold $\theta_2$ (1st row), coupling $\kappa_2$ (2nd row), and RR (3rd row) when fixing agent 1 threshold $\theta_1$ and coupling strength $\kappa_1$. Grey line: Optimal symmetric $\theta$ and $ \kappa$. {\bf c.} Patch departure decision time distributions for each group in ({\bf b}).}
\label{fig3}
\end{figure}

Relaxing assumptions of the above idealized models, we now examine how group strategy parameterization shapes foraging efficiency. In a homogeneous group (individuals in the group have same decision threshold $\theta$ and coupling strength $\kappa$) with diffusive coupling, increasing the coupling strength increases the RR, as long as the decision threshold is properly tuned (along blue line) due to noise cancelation (Fig.~\ref{fig3}a). 

If we break the symmetry of the diffusive coupling model (differing coupling strengths or decision thresholds), individuals may value their neighbor's beliefs less (lower $\kappa_i$) or may need less evidence to decide to depart the patch ($\theta_i$ closer to zero) which could occur due to varying experience or access to information across the group. To identify how asymmetric strategies shape group performance, we consider two agents with different decision thresholds $\theta_1 \neq \theta_2$ and coupling strengths $\kappa_1 \neq \kappa_2$. We compare four representative cases, different combinations of high and low decision thresholds ($\theta_1$) and coupling parameters ($\kappa_1$) for agent 1, in which we optimize the other agent's ($\theta_2$, $\kappa_2$). 
When agent 1's fixed diffusive coupling is strong, the group RR can still be near optimal if agent 2 compensates for agent 1's decision threshold being too large or small (Fig.~\ref{fig3}b). However, when agent 1's coupling is weak, the optimal group RR is substantially suboptimal even when agent 2 compensates for decision threshold mistuning. As with symmetric groups, strong social communication is important for efficient group-level foraging. Strong coupling pushes (pulls) the belief of the first agent with a far (close) decision threshold towards their decision threshold such that the foraging time for both agents is similar (Fig.~\ref{fig3}c, blue/red curves). When agent 1 is weakly coupled to agent 2, the best strategy for the group is for agent 2 to compensate for late/hasty decisions by speeding/slowing their own, but the agents decide at different times and the group's RRs are decreased (Fig.~\ref{fig3}c, purple/green curves).

\subsection{Decisions in pulsatile coupled system}
\begin{figure}
\begin{center}
\includegraphics[width=14cm]{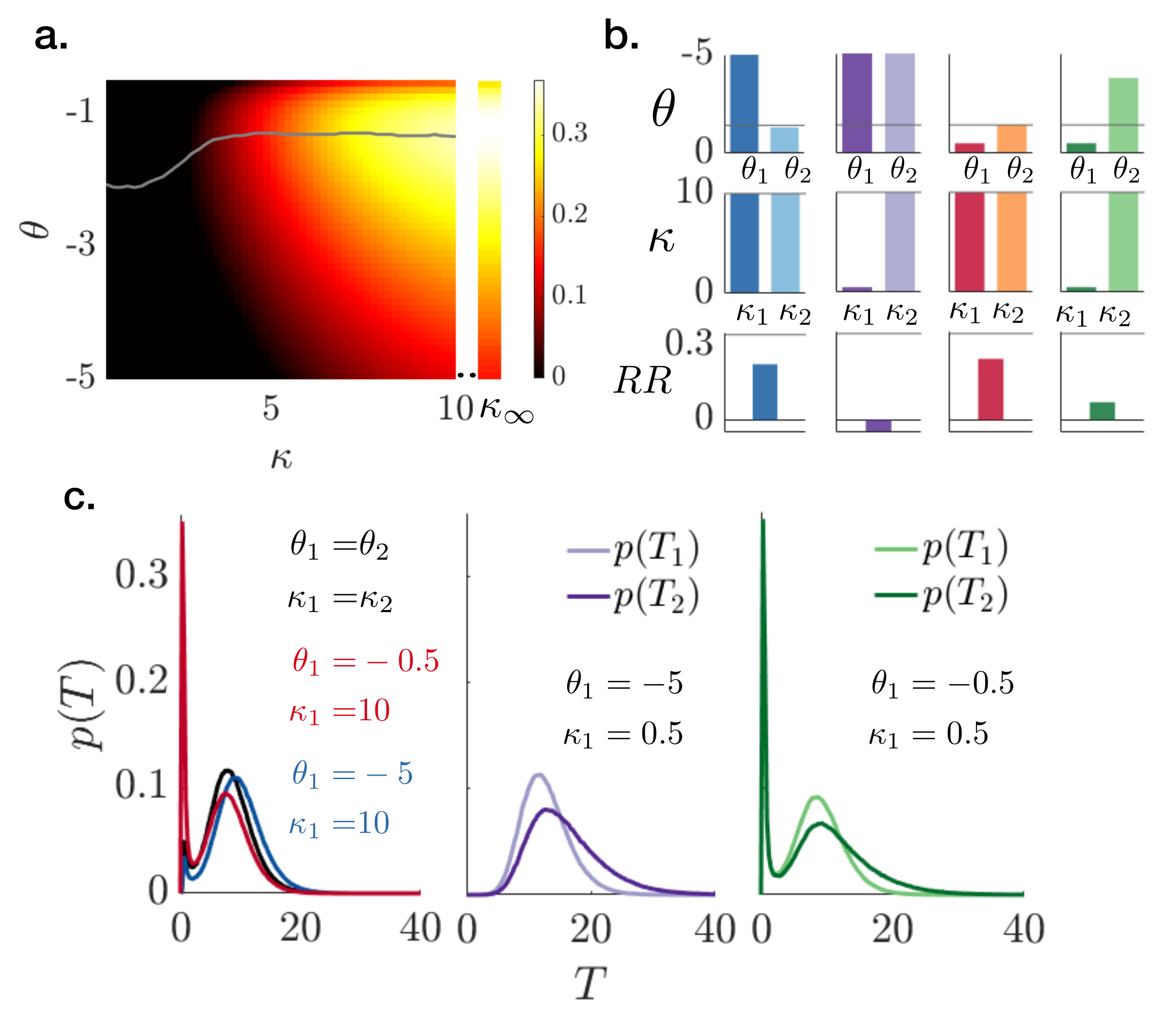}
\end{center}
\vspace{-25pt}
\caption{Comparison of RRs in group with pulsatile coupling. {\bf a.} RR heat map for a two agent group with the same decision threshold $\theta$ and coupling strength $\kappa$, and $\alpha = 1, \rho=4, \tau=5$. Grey line: Optimal decision threshold $\theta$ for a given coupling strength. The line is smoothed by computing average over a sliding window of length $3$. Right column: RR in the infinite coupling limit ($\kappa \to \infty$). {\bf b.} Asymmetric strategies. Optimal threshold $\theta_2$ (1st row), coupling $\kappa_2$ (2nd row), and RR (3rd row) for fixed agent 1 threshold $\theta_1$ and coupling strength $\kappa_1$. Grey line: Optimal symmetric $\theta$ and $\kappa$. {\bf c.} Patch departure decision time distributions for each group in ({\bf b}).}
\label{fig4}
\end{figure}

Pulsatile coupling involves less overall information exchange and yet has a stronger dependence on the coupling parameter, $\kappa$, compared to diffusive coupling (Fig.~\ref{fig4}a).
As with the diffusively coupled group, strong coupling increases groups' RR in a pulsatile coupled system (Fig.~\ref{fig4}a). 
Moreover, the RR falls off sharply for weak coupling (compare Fig.~\ref{fig3}a and Fig.~\ref{fig4}a). For asymmetric strategies, this trend is even apparent when agent 1's coupling is weak but agent 2 employs strong compensatory coupling (Fig.~\ref{fig4}b, purple/green bars). Even with strong bidirectional pulsatile coupling, heterogeneous groups attain lower RRs than heterogeneous diffusively coupled groups (compare Fig.~\ref{fig3}b, blue/red and Fig.~\ref{fig4}b, blue/red). This indicates that while pulsatile coupling strategies can attain high RRs in the symmetric case with intermediate values of $\kappa$, their performance is sensitive to mistuning. Since agents only communicate their decisions to depart, one cannot compensate for the mistiming of their neighbor's departure decisions, but can only synchronize with poor departure decision times to raise the group RR (Fig.~\ref{fig4}c).

\subsection{Model identification and fitting}

\begin{figure}[ht!]
\begin{center}  \includegraphics[width=14cm]{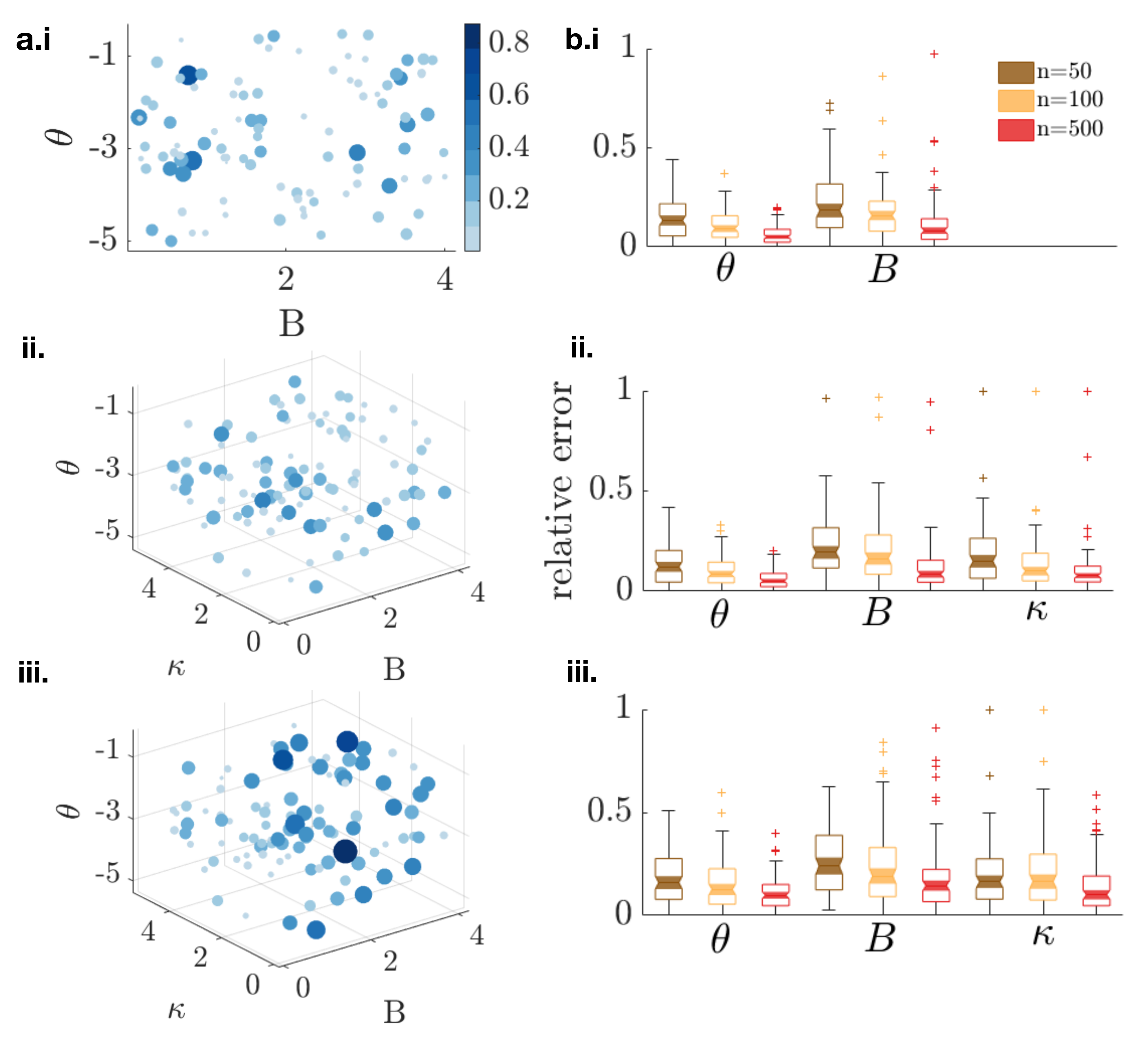}	\end{center}
\vspace{-25pt}
\caption{Parametric fits of no coupling, diffusive, and pulsatile model using Bayesian maximum a posteriori (MAP) estimation. {\bf a.} Estimation errors for {\bf i.}~no information coupling, {\bf ii.}~diffusive coupling, and {\bf iii.}~pulsatile coupling. Bubble location represents true parameters and bubble size/color indicates the total sum of relative estimation error for displayed parameters, Eq.~(\ref{relerr}). Each plot is obtained from $100$ parametric samples from a uniform prior where $\theta \in [-0.1, -5]$, $\kappa \in [0.1,6]$, $B \in [0.1,4]$ given $n=100$ departure time samples each. {\bf b.}  Box plots showing the mean relative error (bars), upper/lower quartiles (boxes), 95\% confidence intervals (error bars), and outliers (stars) in MAP estimate of the model parameters ($\theta$, $B$, $\kappa$) for {\bf i.}~no information coupling, {\bf ii.}~diffusive coupling, and {\bf iii.}~pulsatile coupling models, calculated using data of sample size $n = 50, 100, 500$. The error distribution shown are computed using $100$ parametric samples from a uniform prior.}
\label{fig5}
\end{figure}

To test model identifiability and illustrate model fitting, we developed a method for inferring model parameters from synthetic data (i.e., generated by the model itself) for a group with two agents. Parameters were selected using the Bayesian Maximum a Posteriori (MAP) method~\cite{hastie2001elements}. Using Bayes rule and the independence of each departure time observation pair $\mathbf{T}^k$, we can write down the posterior distribution for the probability of model parameters $\Theta$ given $K$ observed departure decision time pairs $\mathbf{T}^{1:K}$,
\begin{align}   \label{posterior}
     P(\Theta \lvert \mathbf{T}^{1:K}) = \frac{P(\mathbf{T}^{1:K} \lvert \Theta) P(\Theta)}{P(\mathbf{T}^{1:K})} =  \frac{\prod_{k=1}^K P(\mathbf{T}^k \lvert \Theta)}{P(\mathbf{T}^{1:K})} P(\Theta),
\end{align}
where $P(\Theta) = P(\theta) \cdot P(B) \cdot P(\kappa)$ is a jointly independent parametric prior and $P(\mathbf{T^{1:K}})$ is the marginal over the decision time set which can be dropped as it does not depend conditionally on model parameters. The MAP estimate for model parameters $\hat{\Theta}$ is then the mode of the posterior, Eq.~(\ref{posterior}),
\begin{align*}
    \hat{\Theta} & = \argmax_{\Theta} P(\Theta \lvert \mathbf{T}^K) = \displaystyle \argmax_{\Theta} \prod_{k=1}^K P(\mathbf{T}^k \lvert \Theta)P(\Theta),
\end{align*}
selecting $(\theta, \kappa, B)$ for diffusive and pulsatile coupling and $(\theta, B)$ for no coupling. 

We fit models to data from a randomly chosen set of parameters (Fig.~\ref{fig5}a), and compute the error relative to the true parameters used to generate synthetic data,
\begin{align}
    \text{Rel Err} = \frac{1}{3} \left( \frac{\vert \hat{\theta} - \theta^{\rm true} \vert }{\vert \theta^{\rm true} \vert} +  \frac{\vert \hat{B} - B^{\rm true} \vert }{B^{\rm true}} +  \frac{\vert \hat{\kappa} - \kappa^{\rm true} \vert }{\kappa^{\rm true}} \right). \label{relerr}
\end{align}
Parameters for the pulsatile coupled model have the highest relative error, but we find no other systematic variation in error. We conjecture that the higher relative error of the pulsatile coupled model is due to its parameter sensitivity. As such, sampling variability in generating synthetic data makes parameters more difficult to identify.

We also compute average error in parameter estimation for the three different models across the parametric prior (Fig.~\ref{fig5}b). The relative error in the MAP estimate is averaged over $100$ different parameter sets for data of sample size $n = 50, 100,$ and $500$. The average error is slightly higher for the pulsatile model, consistent with the results for individual parameter sets in Fig.~\ref{fig5}a. Generally, increasing the number of decision time pairs sampled tends to decrease the average relative error. One exception is the coupling parameter $\kappa$ in the pulsatile coupling model (Fig.~\ref{fig5}b.iii), which we expect is due to stochasticity in the calculation of the effective average relative error.

\begin{figure}
\begin{center}  \includegraphics[width=6cm]{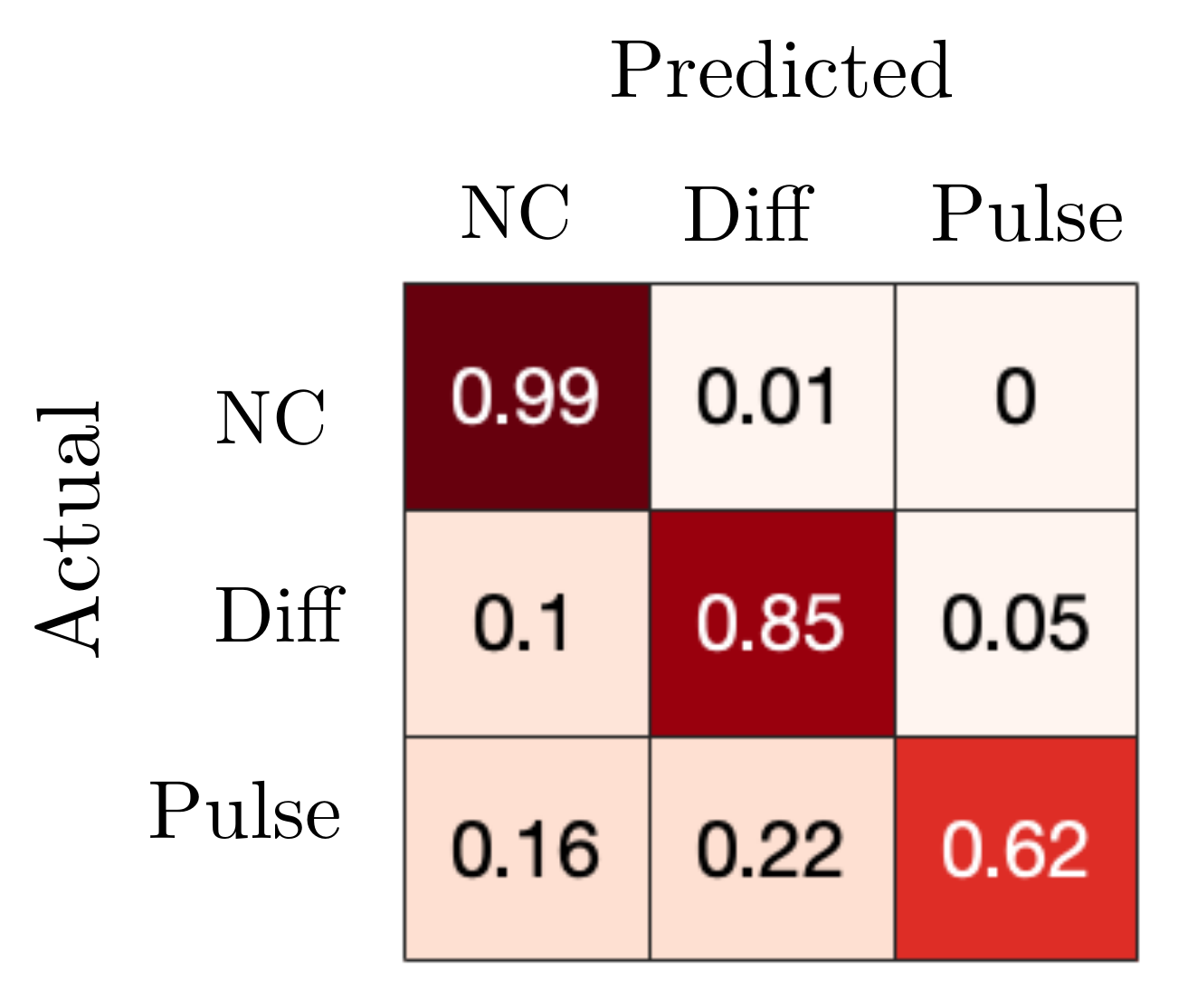}	\end{center}
\vspace{-25pt}
\caption{Confusion  matrix showing model identifiability for no coupling (NC); diffusive coupling (Diff); and pulsatile coupling (Pulse). Numbers and heatmap represent fraction of actual model samples predicted to be each model type, based on a Bayes Factor analysis, Eq.~(\ref{logBF}). Each row is generated from $900$ parametric samples of the actual model using a uniform prior on $\theta \in [-5,-0.1]$, $\kappa \in [0.1,6]$, $B \in [0.1,4]$. Each model fit uses $n=50$ decision time pairs.}
\label{fig6}
\end{figure}

To develop a framework for determining the modes of communication used by foraging animals, we calculate Bayes factors to determine which of two models classes of a pair better explains data. Bayes factors are a model comparison measure, which are ratios of the likelihood of either model class across all possible parameter values~\cite{gelman2013bayesian}.
Considering two such models $\{ M_1, M_2 \} \in \{ M_{\rm no \ coupling}, M_{\rm diff}, M_{\rm pulse} \}$, we can define the log likelihood ratio of either model $M_j$ given a set of observed decision times and then use Bayes rule to rearrange
\begin{align*}
    \log {\rm BF} = \frac{P(M_1 | \mathbf{T}^{1:K})}{P(M_2 | \mathbf{T}^{1:K})} = \frac{P(\mathbf{T}^{1:K} | M_1) P(M_1)}{P( \mathbf{T}^{1:K} | M_2) P(M_2)},
\end{align*}
where $\mathbf{T}^{1:K}$ is the set of $K$ patch departure decision time pairs. For equally likely models $P(M_1) = P(M_2)$, and we can marginalize over parametric priors to obtain:
\begin{align}
     {\rm \log BF} = \log \frac{\int_{\Omega_{\Theta_{M_1}}} P(\mathbf{T}^K \lvert \Theta_{M_1}) P(\Theta_{M_1}) d \Theta_{M_1} }{\int_{\Omega_{\Theta_{M_2}}} P(\mathbf{T}^K \lvert \Theta_{M_2}) P(\Theta_{M_2}) d \Theta_{M_2}}, \label{logBF}
\end{align}
where $\Theta_{M_j}$ is the set of parameters for model $M_j$ with corresponding parametric prior $P(\Theta_{M_j})$. Positive (negative) log Bayes Factors provide evidence for model $M_1$ ($M_2$). 
For a randomly chosen parametric set of an actual model, the predicted model is the one that always generates $\rm \log BF >0$ when in the numerator ($M_1$). Given 900 parametric samples from a uniform prior for each model, we empirically generate a likelihood of each model class being  identified as itself or another model (Fig.~\ref{fig6}). The model without coupling is easily identified whereas the coupled models are misclassified more often. Weak coupled model parameters can lead to misidentification as a non-coupled model. The pulsatile coupled model is the most difficult to identify. We conjecture this is because of its goodness-of-fit sensitivity to the coupling parameters (which is penalized by Bayes factors). The diffusive model, on the other hand, is a more robust model, so it is identified more often, as often predicted when synthetic data from the pulsatile model is provide. Note, these results do not change significantly as the number of decision time samples is increased (See Fig.~\ref{figs1} in Appendix).

We conclude that the pulsatile model can perform well despite only a small amount of information exchange, but does require fine-tuning and can be difficult to identify. The diffusive coupled model is more robust and does require fine tuning, but involves much more information exchange. Our model identification results suggest that we can expect reasonable asymptotic model class identifiability even with only $n=50$ patch departure decision observations for a pair, which suggests that the analysis provided here could translate well to identifying the strategies used by animals in the field.

\section{Discussion}
Many animals forage in groups, but most models focus on the mechanics of single foragers patch departure decision process or greatly simplify the decision formation and belief sharing processes of groups. In this work we extended our previous mechanistic models of patch foraging decisions \cite{davidson19,kilpatrick21} to consider social information sharing among a cohesive foraging group. Group cohesion is formulated as a constraint, and represents the movement dynamics, for example, baboons or capuchin monkeys which tend to move together. Individuals in the group must not only infer the current state of resources to choose an efficient time to leave the patch, but also must synchronize decisions for the group to stay together. Poor synchronization decreases the group reward rate, due to wasted time of early deciders. Information sharing not only helps synchronize the group but also improves the accuracy of noisy or imperfect decision processes.
We considered two different form of information coupling - diffusive (full sharing of an agent's belief about the current patch quality) and pulsatile (sharing only of time of decision to leave). We asked how these couplings affect overall foraging efficiency, and further examined how the coupling parameters and type can be inferred from data.
High information coupling leads to efficient group decisions (i.e.\ maximizing the average reward among group members) through synchronization of departure decisions. Over the parameter ranges we tested, pulsatile coupling can yield similar or higher average return rates than diffusive coupling, despite its much smaller amount of information exchanged, because it more readily promotes group synchronization in decision times. On the other hand, the pulsatile model must be fine tuned, due to it being an intermittent form of communication, and so performance falls off more rapidly than in the diffusive coupled model.

We determined the identifiability of model parameters and data by fitting models back to synthetic data. Both coupling and decision threshold parameters were readily inferred when the model class was known. Although, the pulsatile model's parameters were more difficult to identify, likely due to the model's sensitivity. When the model class was not known, the diffusive coupled model was occasionally misclassified (mostly as a model with no coupling) and the pulsatile model was misclassified more often (typically as a diffusive coupled model).
Model parameter and class identification tended to asymptote when only using 10 to 100 decision time pairs, a feasible volume of  social foraging field data~\cite{egert2018resource,harel2017social,davis2018estimating}.
This approach therefore provides a clear framework for identifying information exchange strategies during social patch foraging, and for fitting models to data.

Social foraging is crucial for animals as it is important for resource localization and collective search ~\cite{cvikel2015bats,egert2018resource,prat2020decision,Strandburg-Peshkin_Farine_Couzin_Crofoot_2015,di2001social,aureli2008social,martinez2013optimizing}.
While we considered cohesive group movement, other animals such as spider monkeys live in groups but do not maintain cohesion during daily foraging, instead leaving and re-joining groups as they forage (so-called fission-fusion group dynamics)~\cite{chapman1995ecological,havmoller2021arboreal}. 
An extension of our modeling approach could be used to represent fission-fusion group dynamics by considering agents that move between patches in a foraging landscape.
Although note that factors other than reward optimization also play a strong role in driving group organization including predator protection, environmental constraints, and mating behavior. Current work seeks to examine evolutionary drivers of differences in group social dynamics within and across species, which could provide a broader class of group performance measures and communication modalities for quantitative models~\cite{strandburg2017habitat,davis2018estimating}.


Our model represents patch-leaving decisions using an accumulation-to-bound process~\cite{davidson19}, where individuals incorporate both personal and social information in order to determine when to leave a patch.
Other work has also used a mathematical formulation of information sharing similar to linear diffusive coupling~\cite{srivastava2014collective}, showing it reduces noise in belief estimates.
It is important to consider various forms of information sharing (e.g. diffusive versus pulsatile coupling) and the way in which such shared information is translated into a decision. Other work has considered nonlinear (e.g., sigmoidal) interactions and information sharing in social groups \cite{zhong_continuous_2019,bizyaeva_patterns_2021}. For larger groups, an alternative representation could treat beliefs as a ``complex contagion," based on a fraction of connected neighbors that have made a decision, instead of a simple sum~\cite{dodds_universal_2004,centola_complex_2007}.
Furthermore, individuals may not be uniformly connected to others so that influential individuals have more weight in their information sharing with neighbors~\cite{conradt2003group,daniels_quantifying_2021,davis2022using}, is an interesting further extension of our model approach, and can help inform how different species exchange information~\cite{janmaat2021using}.


To socially forage, animals often use social cues, informing their decisions through a variety of sensory modalities: the observed harvest or departure of conspecifics from patches~\cite{michelena2009effects}, olfactory cues (e.g., informative breath~\cite{laidre2009informative}); vocalizations~\cite{gillam2007eavesdropping,boinski1993vocal, kohles2020socially}; or visual signals~\cite{fernandez2004visual}. Birds and other species have broad-ranging sensory abilities to detect conspecifics: scavengers can visually detect a conspecific circling a carcass from many kilometers~\cite{harel2017social}, and marine birds can spot diving neighbors~\cite{evans2019social}. Our model forms a basis for future studies that could incorporate more nuanced spatiotemporal features and modalities of communication.

Increasingly sophisticated recording technology facilitates high-resolution motion tracking of diverse species~\cite{westley2018collective,delellis2014collective}, allowing for a thorough validation of theoretical models~\cite{berdahl2018collective}. The availability of such technologies and the gathered large scale data can allow model fitting to foraging behavior; e.g., building on the use of drift-diffusion models fit to evidence accumulation or visual search tasks~\cite{ratcliff2016diffusion}.
By fitting to simulated data, we showed that the parameters in our model can be readily identified, but if the coupling type is unknown, inferring the pulsatile coupling strategy and its parameters is more difficult than inferring diffusive coupling. Future work can build on these results to fit to social foraging data, and infer the information-sharing strategies and differences among individuals.


The social foraging model presented in this study can be extended in multiple ways. First, while some animal groups move as a cohesive whole on the landscape, other groups (for example, spider monkeys) have fission-fusion dynamics. One can relax the assumption that foragers leave together, and represent fission-fusion group dynamics by considering agents that move between patches in a foraging landscape.
Second, one can introduce a variety of biases that animals exhibit into these models such as satisficing or state-dependence (e.g., hunger or thirst) \cite{nonacs2001state}. Third, one can study effects of social structure observed in different animal groups, such as either hierarchical or egalitarian \cite{conradt2003group}, as well as forms of coupling on the foraging dynamics, opening up the opportunity to link collective social structure to collective and individual foraging dynamics.  \\

\noindent
{\bf Code and data accessibility:} Software generating model statistics, identifying models, and plotting figures can be found at \url{https://github.com/sbidari/socialforage}
\section*{Appendix}
\vspace{-5 mm}
\subsection{Departure time statistics calculation}
Decision and departure time statistics are obtained either by running Monte Carlo simulations (on the order of $10^5$ per parameter set) or simulating the corresponding Fokker-Planck equation and calculating the flux through decision boundaries at this $x_1 = \theta_1$ and $x_2 = \theta_2$.
Although the model formulation is general, we focus the analysis in this paper on the tractable case of two coupled agents.
\subsubsection{Monte Carlo simulations}
For a pair of agents, a single Monte Carlo simulation initializes $(x_1(0),x_2(0)) = (0,0)$ and employs the Euler-Maruyama algorithm with the time step $\Delta t =0.01$ for the corresponding pair of coupled Langevin equations until either $x_1(t) \leq \theta_1$ or $x_2(t) \leq \theta_2$. Thereafter, for diffusive coupling the remaining agent's belief $x_j$ evolves with $x_k$ replaced by $\theta_k$ until $x_j \leq \theta_j$. For pulsatile coupling, the remaining agent's belief is instantaneously incremented $x_j(t) - \kappa_j$ and evolves until $x_j \leq \theta_j$. The resulting first and second decision times $T_1$ and $T_2$ are recorded and used to compute statistics.
\subsubsection{Fokker-Planck equation simulations}
When the group's decision time can be described by a single variable system (perfectly coupled cases considered in Section~\ref{perfcoup}), the departure time statistics are computed by numerically simulating the associated Fokker-Planck equations. The Fokker-Planck equation takes the form
\begin{equation} \begin{split} \label{fpe}
    \frac{\partial P(x,t)}{\partial t} = - \frac{\partial}{\partial x} \left( \Big( \rho e^{-\frac{N \, t}{\tau}} - \alpha \Big) P(x,t) \right) + \tilde{B} \frac{\partial^2}{\partial x^2} P(x,t), \\
    P(\theta,t) = 0, \qquad \Big( \rho e^{-\frac{N \, t}{\tau}} - \alpha \Big) P(L,t) = \tilde{B} \frac{\partial}{\partial x} P(L,t)
\end{split} \end{equation}
and is defined over the domain $[\theta, L]$, where $x = \theta$ is an absorbing boundary and $x = L$ ($L$~sufficiently large) is a reflecting boundary. The value of $\tilde{B}$ is $B(B/2)$ for pulsatile (diffusive) coupling. Eq.~(\ref{fpe}) is numerically simulated  using a second order finite difference method in both time and space dimensions with $\Delta t = 0.005$ and $\Delta x = 0.1$ to obtain the probability density for the decision variable $x$, $P(x,t)$. The time dependent probability of departure decisions is obtained by calculating flux through the absorbing boundary $x =\theta$.
\subsection{Likelihood functions calculation}
Likelihood functions $P(\mathbf{T} \lvert \Theta)$ used for model identification and model fitting were obtained by solving the associated Langevin equations using Monte Carlo simulations for a discrete set of parameters which covers the space of the uniform prior. We discretized the parameter space $\Theta$ into a three-dimensional grid $(\theta, B, \kappa) \in [-5,-0.1] \times [0.1,4] \times [0.1,6]$ with 40 partitions along each dimension. Decision time distributions for each parameter set are binned as a probability mass function along $(T_1, T_2) \in [0,75]\times [T_1, 75]$ with steps of size $\Delta T = 0.5$ along each dimension.

\subsection{Model identification with data samples of varying length}
\begin{figure}[t!]
\begin{center}  \includegraphics[width=15cm]{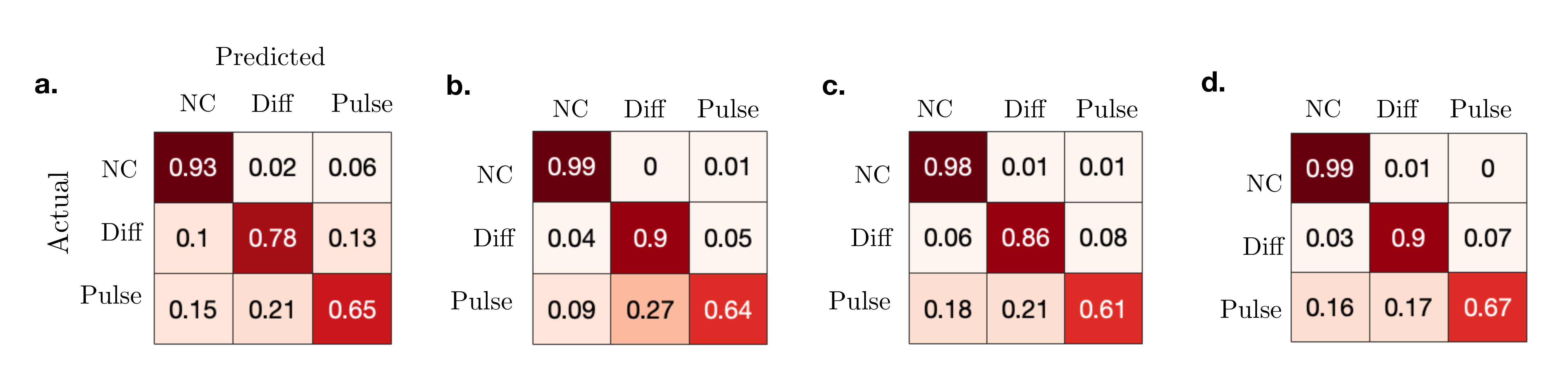}	\end{center} \caption{Confusion  matrix showing identifiability of model classes using data of sample sizes {\bf a.} $n = 10$; {\bf b.} $n = 100$; {\bf c.} $n = 500$; {\bf d.} $n = 1000$. Model parameters are drawn randomly with uniform prior as in Fig.~\ref{fig6}.}
\label{figs1}
\end{figure}

Here show the general trends reported above (Fig.~\ref{fig6}) are preserved when identifying model class using various volumes of decision time sets (Fig.~\ref{figs1}). The model without coupling is easily identified whereas the coupled models are more likely to be misidentified, especially the pulsatile coupled model. Moreover, note there is not substantial improvement in the model class identification fractions when increasing to $n=1000$ samples per model fit.

\end{document}